\documentstyle[12pt]{article}
\input epsf

\begin{document}
\baselineskip=20.5pt

\def\beqra{\begin{eqnarray}} \def\eeqra{\end{eqnarray}}
\def\beqast{\begin{eqnarray*}} \def\eeqast{\end{eqnarray*}}
\def\beq{\begin{equation}}      \def\eeq{\end{equation}}
\def\be{\begin{enumerate}}   \def\ee{\end{enumerate}}

\def\fnote#1#2{\begingroup\def\thefootnote{#1}\footnote{#2}\addtocounter
{footnote}{-1}\endgroup}

\def\ut#1#2{\hfill{UTTG-{#1}-{#2}}}
\def\fl#1#2{\hfill{FERMILAB-PUB-94/{#1}-{#2}}}
\def\itp#1#2{\hfill{NSF-ITP-{#1}-{#2}}}

\def\bet{\beta}
\def\gam{\gamma}
\def\Gam{\Gamma}
\def\la{\lambda}
\def\eps{\epsilon}
\def\La{\Lambda}
\def\si{\sigma}
\def\Si{\Sigma}
\def\al{\alpha}
\def\Tha{\Theta}
\def\tha{\theta}
\def\vphi{\varphi}
\def\del{\delta}
\def\Del{\Delta}
\def\ab{\alpha\beta}
\def\om{\omega}
\def\Om{\Omega}
\def\mn{\mu\nu}
\def\mun{^{\mu}{}_{\nu}}
\def\kap{\kappa}
\def\rsi{\rho\sigma}
\def\beal{\beta\alpha}

\def\til{\tilde}
\def\rta{\rightarrow}
\def\eqv{\equiv}
\def\nab{\nabla}
\def\pa{\partial}
\def\sit{\tilde\sigma}
\def\ul{\underline}
\def\indt{\parindent2.5em}
\def\nd{\noindent}

\def\rsi{\rho\sigma}
\def\beal{\beta\alpha}

\def\caa{{\cal A}}
\def\cb{{\cal B}}
\def\cac{{\cal C}}
\def\cd{{\cal D}}
\def\ce{{\cal E}}
\def\cf{{\cal F}}
\def\cg{{\cal G}}
\def\cah{{\cal H}}
\def\ci{{\cal I}}
\def\cj{{\cal{J}}}
\def\ck{{\cal K}}
\def\cl{{\cal L}}
\def\cm{{\cal M}}
\def\cn{{\cal N}}
\def\cO{{\cal O}}
\def\cp{{\cal P}}
\def\car{{\cal R}}
\def\cs{{\cal S}}
\def\ct{{\cal{T}}}
\def\cu{{\cal{U}}}
\def\cv{{\cal{V}}}
\def\cw{{\cal{W}}}
\def\cx{{\cal{X}}}
\def\cy{{\cal{Y}}}
\def\cz{{\cal{Z}}}

\def\cdgc{C^{\dagger}C}
\def\ccdg{CC^{\dagger}}
\def\pcdgc{C'^{\dagger}C'}
\def\pccdg{C'C'^{\dagger}}
\def\vdgv{v^{\dagger}v}
\def\gmn{\cg^{\mu}_\nu}
\def\gmnb{\cg^{~\,\mu}_{0\,\nu}}
\def\smn{\Sigma^{\mu}_\nu}

\def\raisenot{\raise .5mm\hbox{/}}
\def\nota{\ \hbox{{$a$}\kern-.49em\hbox{/}}}
\def\notA{\hbox{{$A$}\kern-.54em\hbox{\raisenot}}}
\def\notb{\ \hbox{{$b$}\kern-.47em\hbox{/}}}
\def\notB{\ \hbox{{$B$}\kern-.60em\hbox{\raisenot}}}
\def\notc{\ \hbox{{$c$}\kern-.45em\hbox{/}}}
\def\notd{\ \hbox{{$d$}\kern-.53em\hbox{/}}}
\def\notbd{\ \hbox{{$D$}\kern-.61em\hbox{\raisenot}}} 
\def\note{\ \hbox{{$e$}\kern-.47em\hbox{/}}}
\def\notk{\ \hbox{{$k$}\kern-.51em\hbox{/}}}
\def\notp{\ \hbox{{$p$}\kern-.43em\hbox{/}}}
\def\notq{\ \hbox{{$q$}\kern-.47em\hbox{/}}}
\def\notW{\ \hbox{{$W$}\kern-.75em\hbox{\raisenot}}}
\def\notz{\ \hbox{{$Z$}\kern-.61em\hbox{\raisenot}}}
\def\notpa{\hbox{{$\partial$}\kern-.54em\hbox{\raisenot}}}

\def\fo{\hbox{{1}\kern-.25em\hbox{l}}}  
\def\rf#1{$^{#1}$}
\def\bx{\Box}
\def\tr{{\rm Tr}}
\def\rmtr{{\rm tr}}
\def\dgg{\dagger}
\def\trm{{\rm tr}_{_{\left(M\right)}}~}
\def\trn{{\rm tr}_{_{\left(N\right)}}~}
\def\trnn{{\rm tr}_{_{\left(N+1\right)}}~}
\def\tr2n{{\rm tr}_{_{\left(2N\right)}}~}

\def\lag{\langle}
\def\rag{\rangle}
\def\bmid{\big|}

\def\vlap{\overrightarrow{\La p}} 
\def\lrta{\longrightarrow} \def\lrar{\raisebox{.8ex}{$\longrightarrow$}}
\def\rlarw{\longleftarrow\!\!\!\!\!\!\!\!\!\!\!\lrar}
\def\vx{\vec x}  
\def\vy{\vec y}  

\def\llra{\relbar\joinrel\longrightarrow}           
\def\mapright#1{\smash{\mathop{\llra}\limits_{#1}}} 
\def\mapup#1{\smash{\mathop{\llra}\limits^{#1}}} 

\def\nmasymptotic{
{_{\displaystyle{\rm lim}}\atop
{\scriptstyle N,M\rightarrow\infty}
}\,\, 
}

\def\nasymptotic{{_{\stackrel{\displaystyle\longrightarrow}
{N\rightarrow\infty}}\,\, }} 
\def\masymptotic{{_{\stackrel{\displaystyle\longrightarrow}
{M\rightarrow\infty}}\,\, }} 
\def\wasymptotic{{_{\stackrel{\displaystyle\longrightarrow}
{w\rightarrow\infty}}\,\, }} 

\def\asymptext{\raisebox{.6ex}{${_{\stackrel{\displaystyle\longrightarro
w}{x\rightarrow\pm\infty}}\,\, }$}} 
\def\epsilim{{_{\textstyle{\rm lim}}\atop_{\epsilon\rightarrow 0+}\,\, }} 

\def\7#1#2{\mathop{\null#2}\limits^{#1}}        
\def\5#1#2{\mathop{\null#2}\limits_{#1}}        
\def\too#1{\stackrel{#1}{\to}}
\def\tooo#1{\stackrel{#1}{\longleftarrow}}
\def\nout{{\rm in \atop out}}

\def\one{\raisebox{.5ex}{1}}
\def\BM#1{\mbox{\boldmath{$#1$}}}

\def\ltsim{\matrix{<\cr\noalign{\vskip-7pt}\sim\cr}}
\def\gtsim{\matrix{>\cr\noalign{\vskip-7pt}\sim\cr}}
\def\haf{\frac{1}{2}}


\def\place#1#2#3{\vbox to0pt{\kern-\parskip\kern-7pt
                             \kern-#2truein\hbox{\kern#1truein #3}
                             \vss}\nointerlineskip}

\def\illustration #1 by #2 (#3){\vbox to #2{\hrule width #1 height 0pt 
depth
0pt
                                       \vfill\special{illustration #3}}}

\def\scaledillustration #1 by #2 (#3 scaled #4){{\dimen0=#1 \dimen1=#2
           \divide\dimen0 by 1000 \multiply\dimen0 by #4
            \divide\dimen1 by 1000 \multiply\dimen1 by #4
            \illustration \dimen0 by \dimen1 (#3 scaled #4)}}

\def\ON{{\cal O}(N)}
\def\UN{{\cal U}(N)}
\def\bdPh{\mbox{\boldmath{$\dot{\!\Phi}$}}}
\def\bPh{\mbox{\boldmath{$\Phi$}}}
\def\bPhs{\bPh^2}
\def\sef{S_{eff}[\sigma,\pi]}
\def\sigx{\sigma(x)}
\def\pix{\pi(x)}
\def\bph{\mbox{\boldmath{$\phi$}}}
\def\bphs{\bph^2}
\def\ex{\BM{x}}
\def\exs{\ex^2}
\def\xdot{\dot{\!\ex}}
\def\y{\BM{y}}
\def\ys{\y^2}
\def\ydot{\dot{\!\y}}
\def\pat{\pa_t}
\def\pax{\pa_x}
\def\cia{C_{i\alpha}}
\def\cjb{C_{j\beta}}
\def\Gz{G(z)}
\def\log{{\rm log}~}
\def\Re{{\rm Re}~}
\def\Im{{\rm Im}~}
\def\nh{{\rm non-hermitean matrix}}
\def\det{{\rm det}~}
\renewcommand{\thesection}{\arabic{section}}
\renewcommand{\theequation}{\thesection.\arabic{equation}}

\itp{97}{020}

\hfill{hep-th/9703087}\\

\vspace*{.2in}

\begin{center}
{\large\bf NON-HERMITEAN RANDOM MATRIX THEORY:\\
method of hermitean reduction}\end{center}

\begin{center}
{\bf Joshua Feinberg$^{a)}$ \& A. Zee$^{a,b)}$ }
\end{center}
\vskip 2mm
\begin{center}
$^{a)}${Institute for Theoretical Physics}\\
{University of California,\\ Santa Barbara, CA 93106, USA}\\
\vskip 2mm
$^{b)}${Institute for Advanced Study}\\
{Olden Lane, Princeton, NJ 08540, USA}\\
\end{center}
\vskip 3mm
\begin{abstract}
We consider random non-hermitean matrices in the large $N$ limit. The power of analytic function theory cannot be brought to bear directly to analyze non-hermitean random matrices, in contrast to hermitean random matrices. To overcome this difficulty, we show that associated to each ensemble of non-hermitean matrices there is an auxiliary ensemble of random hermitean matrices which can be analyzed by the usual methods. We then extract the 
Green's function and the density of eigenvalues of the non-hermitean ensemble from those of the auxiliary ensemble. We apply this ``method of hermitization" to several examples, and discuss a number of related issues. 
\end{abstract}

\vspace{25pt}
\vfill
\pagebreak

\setcounter{page}{1}
\hspace{-8mm}{\em   ``I found it difficult,...., to keep my mind from wandering \hfill\\
into the magical world of random matrices." \hfill\\}
\vspace{2mm}
{\em F. J. Dyson}\footnote{F.J.~ Dyson, ``Selected Papers of Freeman Dyson with Commentary"\\
(American Mathematical Society, International Press, 1996), p. 39.}\hfill
\section{Introduction}

When Wigner \cite{wigner} first proposed treating the Hamiltonians of
complex systems as
random matrices, he quite naturally took the matrices to be
hermitean (or
real symmetric for time reversal invariant systems) in
accordance with the
fundamental postulates of quantum mechanics. In the last
few years, there
has been increasing interest in considering non-hermitean
random matrices
in the context of various physical problems. (We will use the
term
``non-hermitean" to include ``real but not symmetric.") We
will mention a
few examples: QCD at finite chemical potential\cite{stephanov}, nuclear decay \cite{german}, dissipative systems \cite{german, dissipative}, and neural
networks \cite{neural}. In all these examples, the random non-hermitean matrices were taken from a Gaussian distribution. Recently, Hatano and Nelson \cite{nelson} (see also \cite{efetov}) have used non-hermitean random matrices to study the pinning of
magnetic flux lines in high temperature superconductors.

In this paper, we discuss some aspects of random
non-hermitean matrix theory, hoping to clarify some of the
existing (and growing, see {\em e.g.}\cite{other} for some recent examples) literature, which in the authors's opinion, contains
a number of confusing and confused statements. We develop a method to deal with
non-hermitean matrices, by reducing the non-hermitean problem to an auxiliary hermitean problem. Unlike many examples that appear in the literature, our
formalism is generic and is not limited to Gaussian ensembles. We then
present some specific results. Some of our results may already be known in
the literature. In particular, as we were completing this work, a series of very interesting papers by Zahed and collaborators\cite{stony} studying non-hermitean matrices appeared.

A basic tool in studying hermitean random matrices $\phi$
(henceforth all
matrices will be taken to be $N\times N$ with $N$ tending to
infinity
unless otherwise specified) is the Green's function defined
by
\beq
G(z)=\langle {1\over N} \rmtr {1\over z-\phi}\rangle
\label{gz}
\eeq
where $<\cdots>$ denotes
averaging
over the random distribution from which the matrices
$\phi$ are drawn.
Diagonalizing $\phi$ by a unitary transformation we have
\beqast
G(z)=\langle {1\over N} \sum_{k=1}^N {1\over z-\lambda_k}\rangle
\eeqast
where the $N$ real numbers $\{\lambda_k\}$ are the eigenvalues of $\phi$.
Thus, $\Gz$ is a meromorphic function with poles at the eigenvalues of $\phi$.
In the
large $N$ limit, the poles merge into a cut (or several cuts)
on the real
axis. Powerful theorems from the theory of analytic
functions can then be
brought to bear to the problem of determining $G(z)$
\cite{BIPZ}. All of 
this is well-known. In contrast, for a non-hermitean matrix,
the
eigenvalues invade the complex
plane.  For
example, for non-hermitean matrices $\phi$ generated with
the probability
$P(\phi)={1\over Z}e^{-N \rmtr \phi^{\dagger}\phi}$,  Ginibre determined
long ago \cite{ginibre} that the eigenvalues are uniformly distributed
over a disk of
radius unity in the complex plane. Another simple example is presented in the Appendix. Thus, we lose the
powerful aid of
analytic function theory. To get oriented, consider the
simple example of
the function
\beqast
g(z)=\int\limits_0^{2\pi} {d\theta \over 2\pi}{1\over z - e^{i
\theta} } = \oint {dw\over 2\pi i} {1\over w(z-w)}
\eeqast
which appears to depend only on $z$. In fact, $g(z,z^*) = \theta\left(|z|^2 -1\right)/z$ is clearly
non-analytic, and depends on both $z$ and $z^*$. Speaking mathematically, one would say that this integral defines two functions, one for $|z|>1$ and one for $|z|<1$.

In the theory of hermitean random matrices, a powerful
method consists of the Feynman diagrammatic expansion\cite{bzw} in which
one expands $G(z)=\sum_{k} \langle \rmtr \phi^k \rangle/z^{k+1}$
as a series in
$1/z$, the `` bare quark propagator." This method is
clearly no longer
available in studying non-hermitean random matrices. The
knowledge of
$G(z)$ as a series in $1/z$ can no longer tell us anything
about the behavior $G(z)$ for small $z$, as the simple example above makes clear. The eigenvalues fill a two-dimensional
region rather than a one-dimensional region, as is the case for hermitean matrices.

A renormalization group inspired method used in \cite{french, rg, daz, rectangles} implicitly involves an expansion in $1/z$ and is thus also not available for dealing with non-hermitean matrices without suitable further developments.

\vspace{3mm}
This paper is organized as follows. In Section 2 we review some of the basic formalism. We also make a tentative connection with the theory of quaternions.
Section 3 is the heart of this paper which consists of showing that the Green's functions and density of
eigenvalues of the random non-hermitean matrix may be determined from
those of an auxiliary hermitean matrix. The hermitean auxiliary ensemble
may be analyzed by the usual methods. We refer to this process as the ``Method
of Hermitization". In Section 4 we show how to determine the boundary of the eigenvalue distribution of the complex
matrix without ever determining its density of eigenvalues $\rho(x,y)$. In Section 5 we present some simple applications of the hermitization method. 
In Section 6 we discuss some aspects of ``energy-level" dynamics of
non-hermitean hamiltonians. In particular, we show how the celebrated
phenomenon of level repulsion in the case hermitean hamiltonians changes
when one discusses non-hermitean hamiltonians. We present a simple (yet instructive) example in the Appendix.

\pagebreak

\section{Some Basic Formalism}
\setcounter{equation}{0}

In this section, we will derive the central formula for
studying the eigenvalue distribution of random non-hermitean matrices.
While this
formula is already known in the literature, we hope to
clarify a number of
misleading statements made in its derivation. In order to be
absolutely
clear, we will begin with some almost laughably elementary
facts.

For $z=x+iy$ we define 
\beq\label{pa}
\pa\equiv {\partial\over \partial z} = {1\over 2}
\left({\partial\over \partial x} -i{\partial\over \partial y}\right)
\eeq
so that $\partial z=1$. Clearly, $\partial z^*=0$. (We denote complex conjugation by *.) 
It follows
that $\partial z^{*n}=0~~~(n\geq 0),$ and $\partial~f(z^*)=0$ in a region in which $f$ has a series expansion in powers of $z^*$. As is
well-known, this fails if $f$ does not have a series expansion.
In the
simplest example, we have 
\beq\label{delta}
\partial {1\over z^*}=\pi
\delta(x)\delta(y)\,,
\eeq
which we can easily prove by
integrating the left and
right hand sides over a small square centered at the origin. Similarly, we 
define
\beq\label{bpa}
\pa^{*}\equiv {\pa\over \pa z^*} = {1\over 2} \left({\pa\over\pa x} + i{\pa\over\pa y}\right)\,,
\eeq
so that
$\pa^{*}~z^*=1$. We have, obviously,
$\pa^{*}~(1/z)=\pi \delta(x)\delta(y)$.

Let us now calculate $\partial\pa^{*}~ \log(z z^*)=\partial\pa^{*}~ 
(\log z +\log z^*)$. At this point it is important to keep in mind that
$\pa~ \log z$ is not $1 \over z$ in the cut plane! ($\pa~ \log z$ is $1 \over z$ on the multi-sheeted Riemann surface of course.) Instead,
the correct formula is 
\beq\label{dlog}
\pa ~\log z = {1\over
z}+\pi
\theta(-x) \delta (y)
\eeq
with the standard choice of cutting
$\log z$ along
the negative real axis. Thus, 
\beqra\label{ddbar1}
\partial\pa^{*} (\log z + \log z^*) &=& \pa^{*} (
{1\over z}+\pi
\theta(-x) \delta (y)) + {\rm c.c.}\nonumber\\{}\nonumber\\ &=& \pi \delta(x) \delta (y) + \pi
\delta(x)
\delta (y)-\pi \delta(x) \delta (y)\,,
\eeqra
where the last term
comes from
$(\partial + \pa^{*})~\pi
\theta(-x)
\delta (y) =-\pi \delta(x) \delta (y) $. We finally obtain
\beq\label{ddbar}
\partial\pa^{*} ~\log(z z^*)
=\pi
\delta(x) \delta (y)\,.
\eeq
Notice that if we had glibly used
$\partial~ \log z={1\over z}$, as was sometimes done in the
literature, we
would have missed the last term in (\ref{ddbar1}) and obtained an
erroneous formula
off by a factor of 2 from the correct formula, thus leading to
an endless
stream of apparent paradoxes.

With these basic facts established we are now ready to deal with the average density of eigenvalues of a non-hermitean matrix
$\phi$. We diagonalize the matrix by a similarity
transformation
\beq\label{similarity}
\phi=S^{-1} \Lambda S
\eeq
where $\Lambda$ denotes a
diagonal matrix with elements $\lambda_i$, $i=1,..., N$. Taking the
hermitean
conjugate we have $\phi^{\dagger}=S^{\dagger} \Lambda^*
S^{-1 \dagger}$. By definition, the density 
of eigenvalues is 
\beq\label{rho1}
\rho(x,y)=\langle {1\over N} \sum_i \delta(x-\Re \lambda_i)~ \delta (y- \Im
\lambda_i)\rangle\,.
\eeq
Given the preceding discussion, we obtain from (\ref{ddbar}) that  $\pa^*~\rmtr~(z-\La)^{-1} \equiv \pa^*~\rmtr~(z-\phi)^{-1} = \pi\sum_i \delta(x-\Re\la_i)~ \delta (y-\Im\la_i)$. Thus, using the identity 
\beqast
\det (z-\La)(z^*-\La^*) = \det (z-\phi)(z^*-\phi^\dgg)
\eeqast
we may express $\rho$ in terms of $\phi$ as 
\beq\label{rho}
\rho(x,y) = {1\over\pi} \partial\pa^{*}~\langle {1\over N} \rmtr~\log~(z-\phi)(z^*-\phi^\dgg)\rangle\,.
\eeq
This formula has been known in the literature (see {\em e.g.,} \cite{german}), 
but there a seemingly unnatural small regulator had been introduced to 
make the argument of the logarithm in (\ref{rho}) strictly positive to avoid
the branch point at $0$. Our derivation of (\ref{ddbar}) suggests that that regulator is not really necessary.

Our expression (\ref{rho}) is manifestly symmetric in $z$ and $z^*$. It arises quite naturally 
in calculations involving Grassmann variables\cite{stephanov, german}
where ${\rm det}~(z-\phi)(z^*-\phi^\dgg)$ is the ``fermion determinant".
However, it involves the logarithm inside the average, as well as two derivatives. Thus, unless one has a simple way of calculating the average in (\ref{rho}) ({\em e.g.,} the replica method in the case of Gaussian ensembles \cite{german}), (\ref{rho}) is not a practical expression for $\rho(x,y)$. In most cases it is easier to calculate the Green's function
\beq\label{greens}
G(z,z^*)=\langle {1\over N} \rmtr {1\over z-\phi}\rangle = \langle {1\over N}
\sum_i {1\over (x-x_i) +
i(y-y_i)}\rangle\,,
\eeq
(with $\lambda_i= x_i+iy_i$) than to do the average $\langle {1\over N} \rmtr~\log~(z-\phi)(z^*-\phi^\dgg)\rangle\,.$ Then, from (\ref{rho1}) and from
the complex-conjugate of (\ref{delta}), we have
\beq\label{rho11}
\rho(x,y) = {1\over\pi} \pa^{*}~G(z, z^*)\,,
\eeq
which is a simpler representation of $\rho(x,y)$ than (\ref{rho}).

These two representations for $\rho(x,y)$ and the relation between them can 
be interpreted by recognizing (\ref{rho}) as a two dimensional Poisson equation for the electrostatic potential $-{1\over 2} \langle {1\over N}~ \rmtr~\log~(z-\phi)(z^*-\phi^\dgg)\rangle$ created by the charge density $\rho(x,y)$. This connection between eigenvalue
distributions of complex matrices and two-dimensional electrostatics 
has been long known in the literature\cite{german, neural}. Continuing along this line, consider the Green's function $G(z,z^*)$ in (\ref{greens}).
If we define the electric field ${\vec E} = (\Re G, -\Im G)$ then
from the definition of $G(z,z^*)$ we have 
\beq\label{electric}
{\vec E} (x) = \int d^2 y~ \rho(\vy) {\vx-\vy\over |\vx-\vy|^2}\,
\eeq
and thus (\ref{rho11}) is simply the statement of Gauss' law for this 
electrostatic problem.

We end this section with a speculative remark. As we pointed out earlier,
in the theory of random
hermitean matrices it is well known that it is much easier to work with
$G(z)=\langle {1\over N} \rmtr {1\over z-\phi}\rangle$ rather than with the density of eigenvalues $\rho(\mu)$ directly, since the power of analytic function can be brought to bear on $G(z)$. Of course, one can go from 
$\rho(x)$ to $G(z)$ with the identity 
\beq\label{plemelj}
{1\over\pi}~\Im{1\over x-i\epsilon}=
{1\over\pi}{\epsilon\over x^2+\epsilon^2}\rightarrow \delta(x)\,.
\eeq

Here, for random non-hermitean matrices our formula (\ref{rho}) gives the
density of eigenvalues $\rho(x,y)$ directly, but this formula is awkward
to work with because of the logarithm. Its direct analog in the theory of
random hermitean matrices would be $\rho(x)={\partial\over\partial x}
\langle {1\over N} \rmtr \theta(x-\phi)\rangle$, which would also be
extremely awkward to work with. It would thus be desirable to define a
quantity analogous to $G(z)$ and develop a method for calculating it.
Reasoning along these lines we are led to an attempt to write a function
of a quarternionic variable, in the same way that going from $\rho(x)$ to
$G(z)$ we went from a function of a real variable to a function of a
complex variable. Indeed, the obvious analog of (\ref{plemelj}) is easy to
find, namely, 
\beqra\label{quatplemelj} &&{1\over 2 j}~ \left[ {1\over
(z-j\epsilon )|z-j\epsilon |} + i {1\over (z-j\epsilon)|z-j\epsilon | }~i
\right]\nonumber\\{}\nonumber\\ &&={\epsilon\over |z-j\epsilon |^3}=
{\epsilon\over \left(x^2+y^2+\epsilon^2\right)^{3/2}}\rightarrow
2\pi\delta(x)~\delta(y)\,. 
\eeqra Here $z=x+iy$ is a complex number,
$\epsilon$ is a small positive real number and $\{1,i,j,k\}$ is the
standard basis of the quaternion algebra\footnote{ One may of course write
an equivalent formula with $j$ replaced everywhere by the third quaternion
basis element $k$. Also, $\epsilon$ may be taken complex.}. (The absolute
value of a quarternion is defined by $|a+ib+jc+kd|\equiv
\sqrt{a^2+b^2+c^2+d^2}$.)

Unfortunately, (\ref{quatplemelj}) as it stands, is less useful than (\ref{rho}), because it leads to the quaternionic Green's function $G(q)$
($q$ being a quaternionic variable)
\beq\label{quatG}
G(q) = \langle {1\over N} \sum_i {1\over (q-\lambda_i) | q-\lambda_i |}\rangle
\eeq
which involves the absolute value operation explicitly, and thus cannot be written as a simple trace like (\ref{gz}). Note, however, that if we split $q$ quite generally into $q=z+j w$ ($z,w$ being ordinary complex variables), then 
$G(z+j w)$ is manifestly non-analytic in $z$ as $w\rightarrow 0$, even before
taking the average.

We pose this as an interesting problem, namely to find a useful quaternionic generalization of (\ref{gz}) for random non-hermitean matrices, and to develop 
the analogue of the work of
Br\'ezin et al. \cite{BIPZ} associated with it.
\pagebreak

\section{The Hermitization Method: Reduction to Random Hermitean Matrices} 
\setcounter{equation}{0}
We have emphasized that a straightforward diagrammatic method is not allowed 
for non-hermitean matrices. Somewhat remarkably, we can arrive at a 
diagrammatic method indirectly. We will show that the problem of determining 
the eigenvalue density of random non-hermitean matrices can be reduced to the problem of determining the eigenvalue density of random hermitean matrices, 
for which the diagrammatic method may be applied.

We start with the representation $\rho(x,y) = {1\over\pi} \partial\pa^{*}~\langle {1\over N} \rmtr~\log~(z-\phi)(z^*-\phi^\dgg)\rangle\,$
(namely, Eq. (\ref{rho}).) By standard manipulations we observe that $\langle\trn\log (z-\phi)(z^*-\phi^\dgg)\rangle =\langle\tr2n\log H\rangle - i\pi N^2$, where $H$ is the hermitean $2N\times 2N$ matrix Hamiltonian 
\beqra\label{H}
H=\left(\begin{array}{cc} 0~~~ & \phi-z\\{} & {}\\
\phi^\dgg-z^* & 0\end{array}\right)\,.
\eeqra
Thus, if we can determine
\beq\label{object}
F(\eta; z, z^*) = {1\over 2N} \langle \tr2n\log (\eta - H)\rangle
\eeq
we can determine $\rho(x,y)$.

Consider now the propagator associated with $H$, namely, 
\beqra
\cg^{\mu}_{\nu} (\eta; z, z^*) = \langle\left({1\over \eta-H}\right)^{\mu}_{\nu}\rangle
\label{propagator}
\eeqra
where $\eta$ is a complex variable and the indices $\mu$ and $\nu$ run over 
all possible $2N$ values. Here we followed a common practice and borrowed some terminology from gauge field theory: we may consider $\phi, \phi^{\dgg}$ as 
``gluons" (in zero space-time dimensions), which interact with a $2N$ 
dimensional multiplet of ``quarks"  $\psi^\mu$, with a complex mass matrix (the ``inverse propagator")
\beqra\label{inverseprop}
\cg_0^{-1} =\left(\begin{array}{cc} \eta~~~ & z\\{} & {}\\
z^* & \eta\end{array}\right)
\eeqra
(expressed in terms of its $N\times N$ blocks.)  The crucial point is that 
since $H$ is hermitean, $\gmn $ can be determined by the usual methods of hermitean random matrix theory. In particular, as we already
mentioned, the diagrammatic evaluation of (\ref{propagator}) is essentially 
the expansion of $\gmn$ in powers of $1/\eta$, with interaction vertices $H$. This is a well defined procedure for large $\eta$, and it converges to a function which is analytic
in the complex $\eta$ plane, except for the cut (or cuts) along the real axis which contain the eigenvalues of $H$. After summing this series (and thus determining $\gmn (\eta; z, z^*)$ in closed form), we are allowed to set $\eta\rightarrow 0$ in (\ref{propagator}). Speaking colloquially, we may say that the crucial maneuver here is that while we cannot expand in powers of 
$z$, we can arrange for $z$ to ``hitch a ride" with $\eta$ and at the end of
the ride, throw $\eta$ away.

We can now calculate $\rho(x,y)$ by two different methods, each with its advantages. The first is to simply observe that 
\beqast
-{1\over H} =\left(\begin{array}{cc} 0~~~ & {1\over z^* - \phi^\dgg}\\{} & {}\\
{1\over z- \phi} & 0\end{array}\right)\quad . 
\eeqast
Thus, the quantity ${1\over z- \phi}$ is simply the lower left block 
of $\gmn(\eta=0; z, z^*)$. In other words, once we have $\gmn (\eta; z, z^*)$
we can set $\eta$ to zero and use (\ref{greens}) and (\ref{rho11})
to write
\beq\label{hermitrho}
\rho(x,y) = {1\over N\pi} \pa^* \tr2n \left[\left(\begin{array}{cc} 0~~~ & {\bf 1}_N\\0~~ & 0~\end{array}\right)\cg (0;z, z^*)\right]\,.
\eeq
This observation is the basis for many of the calculations in 
\cite{stony}. We will illustrate this discussion with a simple example below.

An alternative is to take the trace of $\gmn$:
\beq\label{GH}
\cg(\eta; z, z^*) = {1\over 2N} \langle \tr2n~{1\over \eta - H}\rangle
= {\eta\over N}\langle \trn~{1\over \eta^2 - (z^*-\phi^\dgg)(z-\phi)}\rangle\,,
\eeq
from which one can determine (\ref{object}).

We refer to this procedure of calculating
$G(z, z^*) = {1\over N} \langle \rmtr {1\over z-\phi}\rangle$ and $\rho(x,y)$ 
by manipulating the hermitean matrix $H$ as the ``Method of Hermitization".

We now derive a dispersive representation of $F(\eta; z, z^*)$ in terms of 
$\cg (\eta; z, z^*)$. Recall that the eigenvalue density of $H$
is given by the discontinuity of $\cg$ accross the cut, namely,
\beq\label{omega}
\om(\mu; z, z^*) \equiv {1\over 2N} \langle \tr2n~\delta (\mu - H)\rangle= {1\over \pi} \Im \cg(\mu-i\epsilon)_{|_{\epsilon\rightarrow 0+}}\,. 
\eeq
From (\ref{dlog}) we then find
\beq\label{deta}
{\pa\over \pa\eta} F(\eta; z, z^*) = \cg (\eta; z, z^*) + \pi \left[1-\Omega (\Re \eta)\right]\del(\Im \eta)\,,
\eeq
where $\Omega(\Re \eta; z, z^*)$ is the integrated density of eigenvalues of
$H$,
\beq\label{Omega}
\Om (\mu; z, z^*) = \langle{1\over 2N}~\tr2n~ \theta (\mu-H) \rangle\,.
\eeq

As a result of the ``chiral" block structure of $H$, its spectrum consists of pairs of eigenvalues with equal magnitudes and opposite signs. Namely, if $\xi$ is an eigenvalue of $H$, so is $-\xi$.
This symmetry of the spectrum of $H$ around $\mu=\Re \eta = 0$ 
implies $\Omega (0) = {1\over 2}$. Thus, setting $\eta = is$ pure imaginary we obtain
\beq\label{diffeq}
{\pa\over \pa s} F(is; z, z^*) = i\cg (is; z, z^*) + {i\pi\over 2}\del(s)\,.
\eeq
Integrating over $s$ we have
\beq\label{F0}
F(i0+; z, z^*) = F(is; z, z^*) - i\int\limits_{0+}^{s} ds \cg(is; z, z^*)
\eeq
where the lower integration limit avoids the $\del(s)$ piece in (\ref{diffeq}).
From the definition of $F$, we have that $F(\eta; z, z^*)
\rightarrow \log~\eta$, independent of $z$, as $\eta$ tends to infinity. Thus, applying the operator ${2\over\pi}\partial\pa^{*}$ to both
sides of (\ref{F0}), we finally obtain 
\beq\label{newrho}
\rho(x,y)= -{2i\over\pi}\int\limits_{0+}^{\infty} ds \pa\pa^{*} \cg(is; z, z^*)\,.
\eeq

Thus, the density of eigenvalues of the non-hermitean random matrices $\phi$
is determined by an integral over $\cg(\eta; z, z^*)$. The determination of $\cg(\eta; z, z^*)$ involves only the well-known and
more elementary problem of determining the density of eigenvalues of the
hermitean random matrices $H$, which is the essense of our method of hermitization.

In fact, $H$ in (\ref{H}) consists of the sum of of the deterministic
matrix 
\beqra\label{deterministic}
H_0=\left(\begin{array}{cc} 0~~~ & -z\\{} & {}\\
-z^* & 0\end{array}\right)
\eeqra
and the random matrix 
\beqra\label{random}
V=\left(\begin{array}{cc} 0~~~ & \phi\\{} & {}\\
\phi^\dgg & 0\end{array}\right)\,.
\eeqra
Note that the deterministic piece has $N$ eigenvalues $+r=|z|$
and $N$ eigenvalues $-r$. This problem of adding a deterministic matrix and a random matrix has been extensively discussed in the literature
\cite{bzw, pastur, blue, bluez, bhz, zahed}. The Green's function (\ref{GH}) and the density of eigenvalues (\ref{omega}) are related by the standard expression 
\beq\label{dispersion}
\cg(\eta; z, z^*) =\int\limits_{-\infty}^{\infty} d\mu
{\omega(\mu; z, z^*) \over \eta -\mu} =\int\limits_0^{\infty} d\mu
~{\omega(\mu; z, z^*)}~{2\eta \over
\eta^2 -\mu^2}\,.\eeq
Here we used the fact that $\omega(\mu; z, z^*) = \omega(-\mu; z, z^*)$, 
namely, the symmetry of the spectrum of $H$ mentioned above.

Using the ``dispersive representation" (\ref{dispersion}) we can derive a
compact formal expression for the density of eigenvalues of the 
non-hermitean random matrix $\phi$. Inserting (\ref{dispersion}) into 
(\ref{newrho}) we obtain 
\beq\label{dispersiverho1}
\rho(x,y)=-{4\over \pi}\int\limits_0^\infty d\mu  \pa\pa^{*} \omega(\mu; z, z^*) \int\limits_{0+}^{\infty} ds {s\over s^2+\mu^2}\,.
\eeq
Note that the ``$+$" can now be safely removed from the lower limit of the 
$s$ integral. Also, note that while the $s$ integral appears to be
logarithmithically
divergent, it is in fact convergent thanks to the ``sum rule"
$\int\limits_{-\infty}^{\infty} d\mu
\omega(\mu; z, z^*) =1 $. To make this last observation explicit,
let us recall that $\omega(\mu; z, z^*)\equiv {\partial\Omega(\mu; z, z^*)\over\partial \mu}$, where $\Omega (\mu, z, z^*)$ is the integrated density of eigenvalues defined in (\ref{Omega}). We can thus integrate over $\mu$ by parts in (\ref{dispersiverho1}), thereby 
bringing in $\Om (\mu; z, z^*)$. We obtain finally the compact expression 
\beq\label{dispersiverho}
\rho(x,y)=-{4\over \pi}\int\limits_0^\infty d\mu  \pa\pa^{*} {\Omega(\mu; z, z^*) \over \mu}\,.
\eeq

To summarize, we have obtained a formalism (the ``Method of Hermitization")
for reducing the problem of
dealing with random non-hermitean matrices to the well-studied problem of
dealing with random hermitean matrices. Given the non-hermitean matrix
$\phi$, we study the hermitean matrix $H$ instead. By whatever method one prefers, once one has determined the quark propagator $\gmn$ (or its trace,
the Green's function $\cg (\eta;z,z^{*})$), 
one can in principle obtain $\rho(x,y)$ using (\ref{hermitrho}) or (\ref{dispersiverho}). Whether that can be done in practice is of course 
another story.

Note that a ``shadow" of the variable $\eta$ appears in 
Section 3 of the first paper in\cite{stony}, as a small real
regulator $\lambda$. This regulator explicitly ``breaks" what these authors refer to as the ``holomorphic symmetry", by coupling ``quarks" and ``conjugate quarks", in an anlogous manner to the way a small explicit quark
mass breaks the chiral symmetry in QCD. Then $N$ tends to infinity first,
followed by $\lambda\rightarrow 0$. After these limits were taken, in that particular order, one observes that $\langle\rmtr (z-\phi)^{-1}\rangle$
is not a holomorphic function. Since by now $\lambda=0$, these authors refer 
to this phenomenon as ``spontaneous breakdown of holomorphic symmetry".
From the point of the hermitization method, we see that it is rather misleading to treat $\eta\equiv\lambda$ as a small regulator. On the contrary - one
expands in powers of $1/\eta$ for $\eta$ large, and only after summing this
perturbative series for the hermitean matrix $H$, one is allowed to set $\eta=0$.  From this point of view, as well as from the simple example
given in the Introduction, the non-holomorphy of $\langle\rmtr (z-\phi)^{-1}\rangle$ is not more mysterious than the fact that $\phi$ has complex eigenvalues.

It is worth emphasizing that the formalism developed here has nothing to
do with large $N$ as such. The formalism is of course indifferent to the
method one may choose to use to determine $\gmn(\eta;z,z^{*})$. The problem
of determining the Green's function of chiral matrices such as $H$ has
been discussed by numerous authors \cite{rectangles, bhz, bhznpb, rectangles1}. In some cases, results can be obtained for finite $N$ \cite{bhznpb}. 

It is also clear that the formalism here is totally independent of the form of the probability distribution $P(\phi, \phi^\dgg)$. In special cases one can 
say more. For example, suppose that $P(\phi, \phi^{\dagger})$ is
invariant under 
\beq\label{rotinv} 
\phi\rightarrow
e^{i\alpha}\phi\,,\quad\quad \phi^{\dagger} \rightarrow e^{-i\alpha}
\phi^{\dagger}\,. \eeq 
For instance, this holds if $P$ is constructed out
of $\rmtr ~(\phi^{\dagger}\phi)^m$, $(\rmtr~ \phi^{\dagger}\phi)^n$, etc.
Then, clearly, $\langle {1\over N} \rmtr~\log~(z^*-\phi^\dgg)(
z-\phi)\rangle$ is a function of $|z|^2 = r^2$ rather than of
$z$ and $z*$ separately. The formalism just given simplifies somewhat: the propagator $\gmn (\eta;z,z^{*})$ and the
functions $\cg(\eta;z,z^{*})$, $\omega(\mu;z,z^{*})$, and
$\Omega(\mu;z,z^{*})$ do not depend on $z$ and $z^{*}$ separately, but
only on $r=|z|$. Acting on such functions, the operator $\pa\pa^{*}$ in (\ref{dispersiverho}) reduces to
\beq\label{rotinvdensity}
\pa\pa^{*} = {d\over dr^2} r^2 {d\over dr^2}=
{1\over 4r}{d\over dr} r{d\over dr}
\eeq
The density of eigenvalues $\rho(x,y)=\rho(r)$ is obviously 
circularly symmetric. In particular, the
Gaussian problem considered by Ginibre is of this class.

In some problems, one might be interested in the density of zero
eigenvalues. Expanding
$\Omega(\mu;r)=\Omega_0(\mu)+r^2\Omega_1(\mu)+\cdots$ as a series in
$r^2$, and inserting into (\ref{dispersiverho}) we obtain formally a compact expression for
the density of zero eigenvalues $ \rho(0)=\int {d\mu \over \mu}
\Omega_1(\mu)$

It is worth emphasizing again the point we made in the Introduction,
namely that we can no longer expand in powers of $1/z$ and use the diagrammatic
expansion directly. Consider in particular the class of distribution we just
mentioned. Suppose we are allowed to expand in powers of $1\over z$, then
we would compute 
\beqast
\rho(x,y) &=& {1\over\pi}\partial\pa^{*}~\langle {1\over N}
\rmtr~\log~(z-\phi)\rangle +c.c.\nonumber\\
&=&\delta(x)\delta(y)+ {1\over\pi}
\partial\pa^{*}~ {1\over N} \sum_k {\langle\rmtr~\phi^k\rangle
\over z^k} +c.c.= \delta(x)\delta(y)\,,
\eeqast
since in each term $\langle\rmtr~\phi^k\rangle$ vanishes (for $k>0$ ) by the symmetry of the distribution $P(\phi, \phi^\dgg)$. We would have 
concluded erroneously that $\rho(x,y)$ is concentrated in a
$\delta$-function spike at the origin. 

In contrast, it is of course perfectly legitimate to expand in powers of
$1/\eta$ and use the diagrammatic method to compute $\gmn(\eta; z, z^*)$ 
associated with the hermitean matrix problem.

A similar point worth re-emphasizing is that the formula (\ref{rho11}) for $\rho(x,y)$ is also not very useful as it stands unless we have a way of computing $G(z, z^*)$. What we have to do is to compute $\gmn(\eta; z, z^*)$ by a diagrammatic method and then use (\ref{hermitrho}).

Let us illustrate some of this discussion with a simple example, the 
Gaussian problem first worked out by Ginibre\cite{ginibre}, namely the 
density of eigenvalues of non-hermitean
matrices randomly generated according to the normalized probability 
distribution
\beq\label{ginibreprob}
P(\phi)={1\over Z}e^{-N \rmtr \phi^{\dagger}\phi}\,.
\eeq
According to our method of hermitization,
we can reduce this problem to that of obtaining the density of eigenvalues
of the hermitean Hamiltonian $H$ in (\ref{H}), consisting of a deterministic matrix and a hermitean Gaussian random matrix of a chiral form. As remarked
earlier, this belongs to a class of problems long discussed in the
literature\cite{bzw, pastur, blue, bluez, bhz, zahed}. 

Let us first apply the by-now well-known diagrammatic method to calculate
the Green's function $\cg(\eta; z, z^*)\equiv \cg(\eta; r)$ in the large $N$ 
limit. Recall that the quark self-energy matrix $\smn$ and the quark propagator $\gmn$ in (\ref{propagator}) are related by 
\beq\label{selfenergy}
\gmn =\left( {1\over {\bf\cg}_0^{-1} - {\bf\Sigma}}\right)^\mu_\nu\,.
\eeq
Since ${\bf\Sigma}$ is the sum over all one-quark irreducible graphs we have 
the Dyson-Schwinger identity
\beq\label{SD}
\smn = \langle V{\bf\cg} V \rangle^\mu_\nu\,,
\eeq
where $V$ was defined in (\ref{random}). After a simple calculation we find 
from (\ref{ginibreprob}) and (\ref{SD}) that 
\beq\label{Sigma}
\Sigma^\mu_\nu = \cg(\eta; r) \delta^\mu_\nu\,.
\eeq
Thus, from (\ref{selfenergy}) and (\ref{Sigma}) we observe that the diagonal
blocks of $\gmn$ are equal and proportional to the $N$ dimensional unit matrix.
By ``index democracy" this proportionality constant is nothing but
the Green's function $\cg(\eta;r)\equiv \cg$. We thus have 
\beq\label{gmnginibre}
\gmn = {1\over r^2 - (\eta-\cg)^2} \left(\begin{array}{cc} \cg-\eta & z\\{} & {}\\
z^* & \cg-\eta\end{array}\right)\,.
\eeq
Tracing both sides of (\ref{gmnginibre}) we obtain the ``gap equation"
\beq\label{cubic}
\cg^3-2\eta \cg^2+(1+\eta^2-r^2)\cg-\eta=0\,.
\eeq
The relevant solution can be immediately written down using the classic 
Cardano formula. (Of the three roots, the relevant root is the one which goes 
as $1/\eta$ as $\eta$ tends to infinity, as is evident from the definition 
of $\cg(\eta;r)$.) We will not bother to write the expression explicitly.
Then, following (\ref{hermitrho}) we find 
\beq\label{gin}
\rho(x,y) = {1\over \pi}\pa^* {z^*\over r^2 - \cg^2(0;r)}\,.
\eeq

Solving (\ref{cubic}) at $\eta=0$ we find that either $\cg(0;r)=0$ or
$r^2-\cg^2(0;r)=1$. Plugging $\cg=0$ into (\ref{gin})
we obtain an expression proportional to $\delta(x)\delta(y)$ which is nothing but the zero density outside the eigenvalue distribution. In the other case, substituting 
$r^2-\cg^2(0;r)=1$ in (\ref{gin}) we obtain $\pi\rho=\pa^* z^* =1$, which corresponds to a uniform $\rho$ inside the disk of eigenvalues. These two possibilities 
match at $r=1$ which must therefore correspond to the boundary of the 
eigenvalue distribution. Thus, finally
\beq\label{rhoginibre}
\pi\rho_{_{{\rm Ginibre}}}\!(x,y) = \left\{\begin{array}{c} 1\quad ,\quad r<1\\
{}\\0\quad ,\quad r>1\end{array}\right.
\eeq
in agreement with \cite{ginibre}, but the derivation here is considerably simpler.  It is worthwhile mentioning at this point that the matching of these two solutions at $r=1$ arises naturally in a replica saddle-point calculation of $\langle {1\over N} \rmtr~\log~(z^*-\phi^\dgg)(
z-\phi)\rangle \sim F(0;r)$ along the lines of \cite{german}, as we now
describe briefly. Before the replica limit $n\rightarrow 0$ is taken, 
one deals with $\langle \det^n~(z^*-\phi^\dgg)(
z-\phi)\rangle .$ This expression may be considered as the averaged fermion determinant in the fluctuating gluon field $\phi$ when there are $n$ fermion flavors degenerate in mass. The role of fermion mass is played by $r=|z|$. For $r<1$ there are two large $N$ saddle points, call them A and B, that govern the integral over certain replica matrix fields in the expression for the averaged determinant. These saddle points are obviously in $1-1$ correspondence with the 
two solutions of (\ref{cubic}) at $\eta=0$ that we mentioned before. One of these saddle points, say A, dominates the other ({\em i.e.;} its integrand is larger.) However, when $r>1$, the previously dominating saddle-point A disappears, and the determinant is given entirely by saddle-point B. The two expressions for this averaged determinant as well as their first derivatives with respect to $r$ match continuously at $r=1$. In other words, the 
fermion ``free energy" $\log \langle \det^n~(z^*-\phi^\dgg)(
z-\phi)\rangle $ undergoes a second order phase transition at $r=1$ as
the fermion mass is varied.

\pagebreak

\section{Boundaries of the Eigenvalue Distribution}
\setcounter{equation}{0}

In some applications of random non-hermitean matrices, one is primarily
interested in the domains over which the density of eigenvalues is
non-zero. In this section, we will develop a formalism for determining the
boundary of these domains. 

First, let us get oriented. We have already remarked that the deterministic piece in $H$ has $N$ eigenvalues $+r=|z|$ and $N$ eigenvalues $-r$. 
For small $r=|z|$ we would expect the random piece in $H$ to smear these two adjacent delta function spikes in $\om(\mu; z, z^*)$ into one continuous lump. In other words, the density of eigenvalues $\om$ as a function of $\mu$ is non-zero along a single segment. For large $r$, the deterministic part in $H$ will dominate, and we expect $\om$ as a function of $\mu$ to separate into two lumps centered at $\pm r$. On the other hand, in the original non-hermitean matrix problem, as $|z|$ increases, we expect to arrive at the boundary of the domain outside of which $\rho(x, y)$ vanishes.

It is thus natural and intuitive to expect that the criterion for determining the boundary in question is to locate the points in the complex $z$ plane at which $\om(\mu; z, z^*)$ splits into two lumps. Note that since $\om$
is an even function of $\mu$, as mentioned earlier, the split first occurs at $\mu=0$. We can readily formalize this expectation. Consider the positive 
semi-definite hermitean matrix
\beq\label{hmatrix}
h^2 = (z^*-\phi^\dgg)(z-\phi)
\eeq
whose eigenvalues we denote by $h_i^2$. Then from (\ref{omega}) we have
\beq\label{omega0}
\om(\mu; z, z^*) = {1\over 2N}~ \langle\sum_{i=1}^N [\delta (\mu-h_i) + \delta (\mu + h_i) ]\rangle\,.
\eeq
Thus, if $\om(0;z, z^*)$ vanishes, that implies $h$ and hence $z-\phi$ 
(considered averaged over the randomness of course) do not have any zero eigenvalues, or equivalently, $\phi$ does not have any eigenvalue equal to
$z$. This proves our expectation.

How do we determine the boundary of the domains over which $\rho(x,y)$,
the density of eigenvalues of the non-hermitean matrices $\phi$, is
non-zero? To proceed, we will now broaden our horizons and discuss the
problem for an entire class of which the Ginibre problem that concluded the previous section is an example, namely the class in which the Green's 
function $\cg(\eta;\{r\})$ of the
associated hermitean problem is determined by the solution of a polynomial
equation $F(\cg)=0$. In general, as we will see, there can be a whole set of
parameters $\{r\}$; in Ginibre's problem there is only one real variable
$r$. 

Let us decompose 
\beq\label{reimG}
\cg=u+iv
\eeq
into its real and imaginary parts and imagine moving along just below the real axis (that is, we will take $\eta = \mu - i\epsilon$ as in the previous section.) Note that for $\eta$ real, the coefficients in the polynomial 
$F(\cg)$ are real as well. There will be segments over which $v$ is non-zero; these segments are
where the density of eigenvalues of the hermitean matrix $H$ is non-zero.
(Up to an irrelevant factor, $v>0$ is just a convenient notation for what we
called $\omega$.) Let us now imagine moving along
such a segment. We would like to determine the value of $\mu$ (call it
$\mu_c$) at which $v$ vanishes, in other words, the endpoints of the
segment over which $v$ is non-zero.

Let us decompose the equation $F(\cg)=0$ into its real and imaginary parts
and expand for $v$ small:
\beqra\label{reim}
&& F(u)+h(u)v^2+O(v^4)=0\quad\quad {\rm and}\nonumber\\{}\nonumber\\
&& v((F'(u)+O(v^2))=0\,.
\eeqra
Here $h(u)$ is a function whose detailed form does
not concern us, and $F'$ denotes the first derivative of $F$ with respect
to its argument. All these functions $F$, $F'$, and $h$ are of course also
functions of $\mu$ (and of $\{r\}$ which we think of as fixed.)

At $\mu=\mu_c$, $v$ vanishes by definition and hence according to the first equation in (\ref{reim})
$F(u)$ also vanishes, generically as $F(u)\sim (\mu-\mu_c)$. Thus, we
conclude that $v \sim |\mu-\mu_c|^{1\over2}$. (This type of argument is
of course reminscent of the arguments used in Landau in his theory of
phase transition. As a matter of fact, it explains the robustness of
the square root singularity in $\om(\mu)$ near the edges of the 
eigenvalue distribution for many ensembles of hermitean matrices.)

For $\mu$ away from $\mu_c$, $v$ is non zero, and thus we may take out
the factor of $v$ from the second equation in (\ref{reim}) and write $F'(u)+O(v^2)=0$. Letting $\mu$ go to
$\mu_c$, we argue by continuity that $F'(u)$ also vanishes at $\eta_c$.

Should the reader feel somewhat uneasy over the continuity argument, we
can obtain the same conclusion by differentiating the second equation in (\ref{reim}) with respect to
$\mu$. Using $v \sim |\mu-\mu_c|^{1\over2}$ we have that as $\mu$
approaches $\mu_c$, $dv\over d\mu$ tends to infinity while $dv^3\over
d\mu$ tends to zero. We reach the same conclusion above.

An alternative argument is to write the polynomial $F(\cg)=\prod_{i} (\cg-\cg_i)$
in terms of its roots $\cg_i$. Since all coefficients in $F(\cg)$ are real (we assume that $\eta = \mu - i\epsilon$),  the $\cg_i$ are real or appear as complex-conjugate pairs. In particular, the
``physical" root $\cg(\mu; \{r\})$ is generically complex, and thus appears
with its complex-conjugate partner. As $\mu\rightarrow\mu_c$ (from within
the segment along which $v>0$) $v\rightarrow 0$ and thus the two roots
$\cg(\mu; \{r\})$ and $\cg^*(\mu; \{r\})$ collide, that is to say,
become equal (and real, so that $v(\mu_c)=0$). Thus, $F(\cg)$ has generically
a double zero at $\mu=\mu_c$, and so both $F$ and its first derivative $F'$ vanish at that point.

To summarize, to determine the boundary or edge of the density of
eigenvalues of the hermitean random matrix $H$ we solve the two equations
$F(u)=0$ and $F'(u)=0$ to obtain $\mu_c$ as a function of the parameters
$\{r\}$. In general, there will a whole set of $\mu_c$'s.

This method gives us more than what we need to know in the context of the
non-hermitean problem. To determine the boundary of the density of
eigenvalues of the non-hermitean random matrix $\phi$, we are interested
in the point, as we vary $\{r\}$, when one of the $\mu_c$ appears at the
origin. Thus, we finally obtain the equations to determine the boundary in
question, namely

\beqra\label{boundary}
&& F(u)|_{\mu_c=0}=0\quad\quad {\rm and}\nonumber\\{}\nonumber\\
&& F'(u)|_{\mu_c=0}=0\,.
\eeqra

To see how simply this procedure works apply it to Ginibre's problem in
which $F(u)=u^3-2\eta u^2+(1+\eta^2-r^2)u-\eta$. We obtain almost
instantly $r=1$. Indeed, it is known from Ginibre's work, as well as from our derivation in Section 3, that the eigenvalues are distributed over a disk of radius unity.

In all the examples discussed here and below, $F(\cg)$ is a polynomial of odd
degree in $\cg$. In Ginibre's problem $F$ is a cubic, while in the examples
below, $F$ is a quintic. Furthermore, from (\ref{dispersion}) and the chiral structure of
$H$, we know that $\Re \cg(\eta)$ is an odd function of $\eta$. Thus, at
$\eta=\mu=0$ the real part of $\cg$ vanishes. In general, the imaginary part does not vanish of course, but precisely when the density of eigenvalues
$\omega(\mu)$ splits into two lumps, the imaginary part of $\cg$ also
vanishes at $\mu=0$. It is clear from (\ref{dispersion}) and the symmetry of 
$\om(\mu)$ that in the cut complex plane, away from the cut,
$\cg(\eta)$ is an odd function of its argument. Thus, if we write 
$F(\cg)=\sum_k c_k(\eta) \cg^k$, the coefficient
functions $c_j(\eta)$ must be even in $\eta$ for odd $j$, and odd for even
$j$, namely, obey 
\beq\label{cj}
c_j(-\eta) = (-1)^{j+1}c_j(\eta)\,.
\eeq
One can check by inspection of (\ref{cubic}) that this holds in particular for
Ginibre's polynomial.

This suggests a computationally simpler procedure. Given the polynomial
equation $F(\cg)=0$, first set $\eta$ to zero. Now from (\ref{cj})
$c_0(\eta=0) = 0$, and so we should divide $F(\cg)$ by $\cg$ and set $\cg$ to zero (because we wish to identify the points $\{r\}$ where $\mu_c=0$.)
At the risk of being repetitive but for the sake of clarity, let us apply this to Ginibre's polynomial $\cg^3-2\eta
\cg^2+(1+\eta^2-r^2)\cg-\eta=0$. After the first step, we have
$\cg^3+(1-r^2)\cg=0$. After the second step, we have $r=1$ as desired.

The advantage of this procedure is that in the computation of $\cg$ from the equations analogous to (\ref{selfenergy}) and (\ref{Sigma}),
we can from the
beginning set $\eta$ to zero and then compute various quantities to either
zeroth order or first order in $\cg$. This results in enormous
simplifications.

\pagebreak
\section{Further Examples of the Application of the Method of Hermitization}
The development of the method of hermitization in the previous two sections
was formal and generic. In the last part of Section 3 we demonstrated the
method of hermitization by applying it to Ginibre's problem. In this Section 
we provide further examples to the application of this method.

\subsection{Deterministic plus random, both non-hermitean}
Consider a situation in which the fluctuating non-hermitean matrix 
$\phi$ is shifted by a non-hermitean deterministic piece $h_0$. 
For simplicity, we take $h_0$ to be diagonal, and also assume that $\phi$ is drawn from the Gaussian distribution (\ref{ginibreprob}). Then our discussion of Ginibre's problem (Eq. (\ref{selfenergy}) {\em et. seq.}) follows through with the obvious shifts $z\rightarrow z-h_0$ and $z^*\rightarrow z^*-h_0^\dgg$. 
In particular, (\ref{gmnginibre}) now reads
\beq\label{gmndetran}
\gmn = {1\over R^2 - (\eta-\cg)^2}\left(\begin{array}{cc} \cg-\eta  & z -h_0
\\{} & {}\\
z^* -h_0^\dgg & \cg-\eta \end{array}\right)\,,
\eeq
where $R^2$ is the $N\times N$ diagonal matrix $(z-h_0)^\dgg (z-h_0)$. 
Setting $\eta=0$ and tracing both sides of (\ref{gmndetran}), the ``gap-equation" (\ref{cubic}) changes into
\beq\label{gapeq}
\cg=\cg \,{1\over N} \rmtr \,{1\over R^2-\cg^2}\,.
\eeq
Eq. (\ref{gapeq}) has a trivial solution $\cg=0$ and a non-trivial solution which satisfies $(1/N) \rmtr [1/ (R^2-\cg^2)] = 1\,.$ Note that according to the analysis in Section 4, the boundary of the eigenvalue distribution occurs when the non-trivial solution of (\ref{gapeq}) vanishes as well, and is thus given 
by
\beq\label{detranboundary}
{1\over N} \rmtr \,{1\over (z-h_0)^\dgg(z-h_0)} = 1\,.
\eeq
Extracting the lower left block of $\gmn$ and tracing, we find from (\ref{hermitrho}) that 
\beq\label{detranrho}
\rho(x,y) = {1\over N\pi} \pa^* \rmtr {(z-h_0)^\dgg\over  R^2 -\cg^2}\,.
\eeq
Substituting $\cg=0$  into (\ref{detranrho}) we find $\rho(x,y) = (1/N)~\rmtr~\delta^{(2)} (z-h_0)$, which is simply the spectrum of eigenvalues of $h_0$, and it obviously vanishes outside the domain of the eigenvalue distribution of the random matrix. We thus associate the solution $\cg=0$ 
(\ref{gapeq}) with  the region of zero eigenvalue density outside the domain of
eigenvalue distribution, as we already observed in our analysis of Ginibre's 
problem at the end of Section 3. The eigenvalue distribution is bounded
inside the curve (\ref{detranboundary}), and the density of eigenvalues is
given by (\ref{detranrho}) with $\cg$ being the non-trivial solution of
(\ref{gapeq}). It is instructive to work out some simple examples of this 
case.

\subsubsection{Circular deterministic spectrum}
Assume that the eigenvalues of $h_0$ lie equally spaced along a circle of 
radius $a$. In the limit $N\rightarrow\infty$ they form a uniform
continuum. Defining 
\beq\label{I}
I=\oint {d w\over 2\pi i}~{1\over w^2 - {r^2+a^2-\cg^2\over ar} w+ 1}\,,
\eeq
where $w={\rm exp} ~i\theta$ runs along the unit circle, we have for the non-trivial factor of (\ref{gapeq})
\beq\label{gapcircle}
I=-ar\,.
\eeq 
Let us denote the roots of $w^2-[(r^2+a^2-\cg^2)/ar]w +1 =0$ by $w_1$ and $w_2$. Note that the roots of the quadratic equation satisfy $w_1 w_2 = 1$. Thus, one of the roots, call it $w_{in}$, always lies inside the unit circle, and the other, call it $w_{out}$, outside. We obtain $I=\oint{dw\over 2\pi i} {1\over (w-w_1)(w-w_2)} = {1\over w_{in}-w_{out}}$. Then (\ref{gapcircle}) reduces to $(a^2+r^2-\cg^2)^2-4a^2r^2=1$. There are obviously two roots for $\cg^2$, but one of them is positive for all values of $r$, and thus cannot correspond
to the eigenvalue distribution. The remaining (physical) root is
\beq\label{cgcircle}
\cg^2=r^2+a^2-\sqrt{1+4a^2r^2}\,.
\eeq
The boundary of the eigenvalue distribution occurs at $\cg=0$, namely
at 
\beq\label{annulus}
r^2=a^2\pm 1\,.
\eeq
Thus, if $a>1$, the original circle of eigenvalues
swells into the annulus $\sqrt{a^2-1}<r<\sqrt{a^2+1}$, while for $a<1$, the
eigenvalues fill a disk of radius $\sqrt{a^2+1}$. 
Finally, from (\ref{detranrho}) we find the density inside the annulus (or the
disk) as
\beq\label{circlerho}
\rho(x,y) = {1\over \pi}~\left(1-{a^2\over \sqrt{1+4a^2r^2}}\right)\,.
\eeq
Note that there are more eigenvalues near the outer edge. For the annulus case ($a>1$) the densities at the outer and inner edges are given
by $\pi\rho ({\rm outer}) = 1 - a^2/(2a^2+1)$ and $\pi\rho ({\rm inner}) = 1 - a^2/(2a^2-1)$. For the disk case ($a<1$) the density at the origin is given by $\pi\rho(0) = 1-a^2$ and is thus reduced from the density in Ginibre's case by a factor of $1-a^2$.

In this example the eigenvalues of the deterministic piece $h_0$ were located
along a circle. It would be interesting to generilize this example to 
situations where the eigenvalues of $h_0$ are located along an
arbitrary curve in the complex plane.

\subsubsection{Two point deterministic spectrum}
In our next example we consider an $h_0$ with eigenvalues $\pm\epsilon$,
equally degenerated. With no loss of generality we may always take $\epsilon$ real and positive. The non-trivial factor of (\ref{gapeq}) is now
\beq\label{gap2}
{1\over |z-\epsilon|^2-\cg^2} +  {1\over |z+\epsilon|^2-\cg^2} =2\,.
\eeq
Setting $\cg=0$ in (\ref{gap2}) we conclude that the boundary is determined by 
\beq\label{boundary2}
y^4 + (2x^2 + 2\epsilon^2 -1) y^2 + x^4 - (2\epsilon^2+1)x^2 + \epsilon^4-\epsilon^2 =0\,.
\eeq
Clearly, the eigenvalue
distribution is an even function in both $x$ and $y$. For $\epsilon=0$ this is simply Ginibre's problem and the eigenvalues are distributed uniformly on a disk of unit radius centered around the origin. As we increase $\epsilon$ from zero, 
the disk is distorted: it gets squeezed along the $y$ axis and stretched
along the $x$ axis (in both directions). We find that for $\epsilon<1$ the eigenvalues are still concentrated in a single blob centered at the origin and confined within $x^2<(2\epsilon^2+1+\sqrt{1+8\epsilon^2})/2$. However,
at $\epsilon=1$, what used to be the disk is now completely pinched and separates into two identical lobes that touch only at the origin. As we 
increase $\epsilon$ above $1$, these two lobes separate and their extent along the $x$ axis is determined by $(2\epsilon^2+1-\sqrt{1+8\epsilon^2})/2 <x^2 <(2\epsilon^2+1+\sqrt{1+8\epsilon^2})/2$.  The expression for $\rho(x,y)$, 
(Eq. (\ref{detranrho}))  is not particularly illuminating and we do not
bother to write it here. More generally, if $h_0$ has $K$ different eigenvalues
$\epsilon_i$, Ginibre's disk would get torn into $K$ different blobs as we 
scale $h_0$ up, obviously.

\subsection{Two point deterministic spectrum plus hermitean random}
As a complication of the previous subsection, we will now discuss 
the problem in which $\phi=h_0+M$ consists of the sum of a deterministic
piece $h_0$ and an hermitean random matrix $M$ which we will take from
the Gaussian distribution $P(M)={1\over Z}e^{-{N\over 2}~\rmtr~M^2}$. 
This case is more complicated than the previous cases because the hermiticity 
of $M$ amounts to a constraint on the ``gluon propagator". Our interest in this problem is partly motivated by the flux line pinning problem to be discussed in Section 6. In that problem the eigenvalues of $h_0$ trace out an ellipse. Here we content ourselves with the much simpler case of $h_0$ being a diagonal matrix, with half its eigenvalues equal to $i\epsilon$ and the other half equal to $-i\epsilon$ (with $\epsilon$ real and positive.) The qualitative features of the eigenvalue distribution is clear. For $\epsilon=0$, the eigenvalues are restricted to the real axis and follow Wigner's semi-circle distribution. As $\epsilon$ increases, the eigenvalues invade the complex plane, and eventually, as $\epsilon$ increases past some critical value $\epsilon_c$ we expect the domain over which the density of eigenvalues is non-zero to break up into two ``blobs" centered around $\pm i \epsilon$. Note the obvious symmetries of the shape of the 
eigenvalue distribution: if $(x,y)$ is a point on the boundary, so are
$(\pm x, \pm y)$. Here we shall content ourselves with calculating
the boundary of the eigenvalue distribution only. Due to the specific structure of $h_0$ we break $M$ into $(N/2)\times (N/2)$ blocks
\beq\label{M}
M = \left(\begin{array}{cc} A & C \\{} & {}\\
C^\dgg & B\end{array}\right)\,,
\eeq
in terms of which the Gaussian distribution reads
\beq\label{gaussM}
P(M)={1\over Z}e^{-{N\over 2}~\rmtr~(A^2 +B^2) - N~\rmtr~\cdgc}\,.
\eeq
(In general, of course, we can consider taking the Gaussian widths of $A$, $B$, and $C$ to be different, but we will not entertain this possibility here.)
With $\phi$ and $M$ given above,
the deterministic piece of the Hamiltonian (\ref{H}) in terms of its 
$(N/2)\times (N/2)$ blocks is  
\beq\label{Mdeter}
H_0\equiv -\cg_0^{-1}(\eta=0) = \left(\begin{array}{cccc} 0 & 0 & i\epsilon-z & 0 \\{} & {} & {} & {} \\
0 & 0 & 0 & -i\epsilon-z \\{} & {} & {} & {} \\ -i\epsilon-z^* & 0 & 0 & 0\\
{} & {} & {} & {} \\0 & i\epsilon-z^* & 0 & 0\end{array}\right)\,,
\eeq
and the random part is 
\beq\label{Mrand}
V=\left(\begin{array}{cccc} 0 & 0 & A & C\\{} & {} & {} & {} \\
0 & 0 & C^\dgg & B \\{} & {} & {} & {} \\ A & C & 0 & 0\\
{} & {} & {} & {} \\ C^\dgg & B & 0 & 0\end{array}\right)\,.
\eeq
Speaking picturesquely, and adding to the ``quark-gluon" language mentioned earlier, we can refer to the (block) four dimensional basis as ``flavor", with flavor assignments
\beq\label{quarks}
\psi=\left(\begin{array}{c} u\\c\\d\\s\end{array}\right)\,,
\eeq
containing two ``generations", $(u,d)$ and $(c,s)$.
With this flavor assignment, we see from the interaction $\psi^\dgg V \psi$
that the gluons corresponding to $A$, $B$ and $C$ all act like weak 
interactions gauge bosons. $A$ and $B$ act within their appropriate generation, while $C$ and $C^\dgg$ communicate between the two generations.\footnote{To see this more clearly, and also for computational simplicity, it 
is convenient to change the basis by interchanging rows as well as columns 2 and 3. In the new basis the $A$'s in $V$ will migrate to the upper left
corner, the $B's$ to the lower right corner, the $C$'s to the upper right block, and the $C^\dgg$'s to the lower left block, such that in the new basis $V$ will become simply $M\otimes \sigma_1$. However, we do not use the transformed basis in the text explicitly, to avoid unnecessary repetitions of formulas.} 
Since we are only interested in determining the boundary of the eigenvalue distribution, from now on, we set $\eta=0$, so that $\gmn (\eta=0)$ is hermitean. From (\ref{SD}) and (\ref{gaussM}) we then
find that the self-energy matrix has the form
\beq\label{Mself}
{\bf \Sigma} =\left(\begin{array}{cccc} g_{33}+g_{44} & 0 & g_{31}+g_{42} & 0\\{} & {} & {} & {} \\
0 & g_{33}+g_{44} & 0 & g_{31}+g_{42} \\{} & {} & {} & {} \\ g_{31}^*+g_{42}^* & 0 & g_{11}+g_{22} & 0\\
{} & {} & {} & {} \\ 0 & g_{31}^*+g_{42}^* & 0 & g_{11}+g_{22}\end{array}\right)\,,
\eeq
where the non-zero blocks are proportional to ${\bf 1}_{N\over 2}$ with proportionality
coefficients  that are sums of 
\beq\label{gab}
g_{ab} = g_{ba}^* = {1\over N}~\rmtr (\cg)_{ab~ {\rm block}}\,,\quad a,b =1, \cdots 4\,,
\eeq
where the trace is carried over the $(N/2)\times (N/2)$ blocks. By ``index democracy" and the texture of $H$, we obviously have $g_{11}=g_{22}$ and $g_{33}=g_{44}$. Then clearly $\cg = (g_{11} + g_{33})/2$. Substituting (\ref{Mself}) into (\ref{selfenergy}) we obtain a system of coupled ``gap equations" for the coefficients $g_{ab}$. 
We are interested only in the $g_{ab}$ that appear explicitly
in $\Sigma$.\footnote{Note also that the
trace of the lower left block of $\gmn$ is $g_{31}+g_{42}$. Thus
the $g_{ab}$ that do not appear in (\ref{Mself}) are also irrelevant
for the purpose of calculating $\rho(x,y)$ from (\ref{rho11}).} 
Introducing the quantity $\cd\equiv  (g_{11} - g_{33})/2$ we obtain 
the equations for these elements as
\beqra\label{gapM}
\cd &=& \cg (|g_{13}|^2-|g_{24}|^2 - 4\cg\cd)\nonumber\\
\cg &=& \cg [|g_{13}|^2 + |g_{24}|^2 - 2(\cg^2 + \cd^2)]\nonumber\\
g_{13} &=& (|g_{13}|^2-\cg^2)(z-i\epsilon -g_{13}^*-g_{24}^*)\nonumber\\ 
g_{24} &=& (|g_{24}|^2-\cg^2)(z+i\epsilon -g_{13}^*-g_{24}^*)\,.
\eeqra                        
These are 6 real equations in 2 complex plus 2 real unknowns.
Being interested only in determining the boundary of the eigenvalue distribution, we only need the non-trivial factor of the gap equations to first order in $\cg(0)$. From the first equation in (\ref{gapM}) we find that $\cd=0$ when $\cg=0$ (for the trivial, as well as for the non-trivial factor of the gap equation), thus eliminating one
real unknown and one real equation. Then, dividing the second equation in (\ref{gapM}) by $\cg$ and setting $\cg=0$ we obtain
\beqra\label{gapMM}
 |g_{13}|^2 + |g_{24}|^2 &=& 1\nonumber\\
(z^*+i\epsilon -g_{13}-g_{24})g_{13} &=& 1\nonumber\\ 
(z^*-i\epsilon -g_{13}-g_{24})g_{24} &=& 1\,,
\eeqra            
which constitute 5 real equations in 2 complex unknowns, thus producing a 
single real constraint, namely, the desired equation for the boundary. A direct elimination of one of the $g's$ in (\ref{gapMM}) produces a cubic equation in the other $g$, which has to be solved explicitly. Then, the correct root of 
that cubic has to be fed into the constraint to obtain the boundary. We will
content ourselves with studying perturbatively the two limits
$\epsilon <<1$ and $\epsilon>>1$.

We first examine the small $\epsilon$ limit. It is useful to introduce the combinations 
\beq\label{combo1}
g_{13}+g_{24}=g\sqrt{2}\,,\quad\quad g_{13}-g_{24}=f\sqrt{2}
\eeq
and the parametrizations
\beq\label{para1}
z^*=2\sqrt{2}~{\rm sin}\beta\,,\quad\quad i\epsilon=\lambda\sqrt{2}\,.
\eeq
Then the appropriate linear combinations of the equations (\ref{gapMM}) become
\beqra\label{gapMMM}
&& g^2-2g~{\rm sin}~\beta-\lambda f + 1 = 0\nonumber\\
&& gf -2f~{\rm sin}~\beta-\lambda g = 0\nonumber\\
&& |f|^2 + |g|^2 =1\,.
\eeqra
At $\epsilon=0$ we should recover the ordinary Wigner semi-circular
behavior associated with (\ref{gaussM})
\beq\label{semicircle1}
g_{13}^{(0)}=g_{24}^{(0)} = {z^*-\sqrt{z^{*2}-8}\over 4}
\eeq
namely, 
\beq\label{semicircle2}
g^{(0)} = {z^*-\sqrt{z^{*2}-8}\over 2\sqrt{2}} = -i e^{i\beta}\,,\quad\quad f^{(0)}=0\,.
\eeq
Thus, in a systematic expansion of (\ref{gapMMM}) in powers of $\lambda$ around $\lambda =0$, $f$ is always a small quantity, which is the reason for introducing the combinations (\ref{combo1}). The eigenvalues lie on the real axis in the 
segment $|x|\leq 2\sqrt{2}$ which is the (degenerate) boundary of the 
eigenvalue distribution. In particular, along this ``boundary", the angle
$\beta = \beta^{(0)}\,\quad (0\leq\beta^{(0)}<2\pi)~~$ is real, of course, in accordance with the third equation in 
(\ref{gapMMM}). Let us now solve the first two equations in (\ref{gapMMM}) to leading order in $\lambda$. We find
\beq\label{leadingorder}
g=ie^{i\beta} \left (-1 + {\lambda^2~e^{i\beta}\over 2~{\rm cos}~\beta}\right)\,,\quad\quad f=\lambda e^{2i\beta}\,.
\eeq
Substituting these quantities into the third equation in (\ref{gapMMM})
we find that to this order 
\beq\label{beta}
\beta = \beta^{(0)} -i\lambda^2\,,
\eeq
and thus, we find from (\ref{para1}) that to this order the boundary is \beq\label{boundary1}
x^2 + {4 y^2\over \epsilon^4} = 8\,,
\eeq
which shows that the original segment containing the eigenvalues of $M$ 
swells into an ellipse, with a semi-minor axis
equal to $2\sqrt{2}|\lambda|^2 = \sqrt{2}\epsilon^2$. 

We now turn to the other extreme limit, $\epsilon >>1$. Here we expect of course
that the eigenvalues concentrate in two semicircular distributions around
$z=\pm i\epsilon$. For concreteness, let us concentrate on the vicinity
of $z=-i\epsilon$. Shifting $z^* = i\epsilon +\zeta$, we find that the last two equations in (\ref{gapMM}) become
\beq\label{gapMMMM}
g_{24} + {1\over g_{24}}  = \zeta -g_{13}\,,\quad\quad (2i\epsilon +  {1\over g_{24}}) g_{13} =1\,.
\eeq
We now expand (\ref{gapMMMM}) in powers of $1/\epsilon$, for $|\zeta|/\epsilon <<1$. It is clear that here $g_{13}$ is the small quantity.
To leading order in $1/\epsilon$ we find the semi-circular behavior
\beq\label{semicircle3}
g_{24}^{(0)} = {\zeta - \sqrt{\zeta^2 - 4}\over 2}\,,\quad\quad g_{13}^{(1)} = {1\over 2i\epsilon}\,.
\eeq
Thus, to this order in $1/\epsilon$, the eigenvalues follow semi-circular
distribution along the segments $|x|\leq 2\,, y=\pm \epsilon$, which are
also the (degenerate) boundaries of the eigenvalue distribution.
Note that the width squared of this semi-circle is half that of the semi-circle
in (\ref{semicircle2}), because the eigenvalue distribution is now split into two identical blobs.  It is natural to parametrize
\beq\label{para2}
\zeta=2~{\rm sin}~\gamma\,,
\eeq
in terms of which $g_{24}^{(0)} = -ie^{i\gam}$. Since $\zeta$ is real along the ``boundaries" to leading order in $1/\epsilon$, $\gamma=\gamma^{(0)}\,\quad (0\leq\gamma^{(0)}<2\pi)~~$ is real 
as well, in accordance with the first equation in (\ref{gapMM}). 
Expanding $g_{24}$ to the next order in $1/\epsilon$ we find
\beq\label{leadingorder1}
g_{24} = -ie^{i\gam} \left(1- {1\over 4\epsilon {\rm  cos}~\gam}\right)\,.
\eeq
Substituting these expressions for $g_{24}$ and $g_{13}$ into the first equation in (\ref{gapMM}) we find to this accuracy that on the boundary
\beq\label{gamma}
\gamma= \gam^{(0)} - {i\over 4\epsilon ~{\rm cos}~\gam^{(0)}}\,,
\eeq
and thus from (\ref{para2}) the (lower half of the) boundary is given parametrically by 
\beq\label{boundary2}
\left\{\begin{array}{c} x=2~{\rm sin}~\gam^{(0)}\\
y=-\epsilon + {1\over 2\epsilon}\end{array}\right.
\eeq
We see that the effect of the first non-trivial correction is simply
to pull the two semi-circle distributions towards the real axis in an amount proportional to $1/\epsilon$, with no distortions. We shall present a simple
explanation of this phenomenon in Section 6.

We close this subsection by determining the critical value $\epsilon_c$ at 
which the connected eigenvalue distribution corresponding to small values
of $\epsilon$ breaks into two separate lumps. From the symmetries
of the eigenvalue distribution it is clear that when this happens, the 
boundary passes throught the origin (it then has the shape of the figure 8).
Thus, substituting $z=\beta =0$ in (\ref{gapMMM}) we find a non-trivial solution $g=0, f=1/\lambda$, and thus $\epsilon_c=\sqrt{2}$.

\subsection{QCD at finite chemical potential}
As was mentioned in the Introduction, Stephanov\cite{stephanov} has recently studied the chiral phase transition in QCD at finite chemical potential
using non-hermitean random matrices.
The Euclidean Dirac operator at chemical potential $\mu$ is non-hermitean. The matrix representation of this operator (for massless quarks) is 
\beq\label{diracop}
\phi = \left(\begin{array}{cc} 0~~~ & C + i\mu\\{} & {}\\
C^\dgg + i\mu & 0\end{array}\right)\,.
\eeq
The chiral components of the Dirac operator are represented here by the
$N\times N$ complex matrices $C$ and $C^\dgg$. The fluctuations of the gluonic
background are modeled by taking $C$ to be Gaussian random, namely, with probability distribution $P(C) = {\rm exp} (-N\rmtr \cdgc)/Z$. The Green's function and density of eigenvalues of (\ref{diracop}) were calculated in \cite{stephanov} using the replica method, and were rederived in \cite{stony} diagrammatically. Thus, as a complementary calculation, in the following we shall apply the methods of Section 4 to calculate the boundary of the 
eigenvalue distribution for this ensemble, 
without calculating the density itself. Recall that to the sole
purpose of determining the boundary, we only need the the non-trivial factor of the gap equation to first order in $\cg$ at $\eta=0$. Thus, as in previous sections, we set $\eta=0$
from now on. With $\phi$ given by (\ref{diracop}),
the deterministic piece of the Hamiltonian (\ref{H}) in terms of its 
$N\times N$ blocks is  
\beq\label{diracdeter}
H_0\equiv -\cg_0^{-1}(\eta=0) = \left(\begin{array}{cccc} 0 & 0 & -z & i\mu\\{} & {} & {} & {} \\
0 & 0 & i\mu & -z \\{} & {} & {} & {} \\-z^* & -i\mu & 0 & 0\\
{} & {} & {} & {} \\-i\mu & -z^* & 0 & 0\end{array}\right)\,,
\eeq
and the random part is 
\beq\label{diracrand}
V=\left(\begin{array}{cccc} 0 & 0 & 0 & C\\{} & {} & {} & {} \\
0 & 0 & C^\dgg & 0 \\{} & {} & {} & {} \\ 0 & C & 0 & 0\\
{} & {} & {} & {} \\ C^\dgg & 0 & 0 & 0\end{array}\right)\,.
\eeq
Then from (\ref{SD}) we find\footnote{See footnote 3. Here of course, we do not
have the blocks $A$ and $B$.} that the self-energy matrix has the form
\beq\label{diracself}
{\bf \Sigma} =\left(\begin{array}{cccc} g_{44} & 0 & g_{42} & 0\\{} & {} & {} & {} \\
0 & g_{33} & 0 & g_{31} \\{} & {} & {} & {} \\ g_{42}^* & 0 & g_{22} & 0\\
{} & {} & {} & {} \\ 0 & g_{31}^* & 0 & g_{11}\end{array}\right)\,,
\eeq
where the non-zero blocks are proportional to ${\bf 1}_N$ with proportionality
coefficients  
\beq\label{gab}
g_{ab} = g_{ba}^* = {1\over N}~\trn(\cg)_{ab~ {\rm block}}\,,\quad a,b =1, \cdots 4\,.
\eeq
By ``index democracy" all the diagonal coefficients $g_{kk}$ are equal, and are therefore equal to $\cg$. Similarly, $g_{42}=g_{31}\equiv G$ (they are in the same block of $\Sigma$.)  Substituting (\ref{diracself}) into (\ref{selfenergy}) we obtain a system of coupled ``gap equations" for the coefficients $g_{ab}$. 
We are interested only in the $g_{ab}$ that appear explicitly
in $\Sigma$.\footnote{Note also that the
trace of the lower left block of $\gmn$ is $(g_{31}+g_{42})/2 \equiv G$. Thus
the $g_{ab}$ that do not appear in (\ref{diracself}) are also irrelevant
for the purpose of calculating $\rho(x,y)$ from (\ref{rho11}).} Among the gap
equations there is subset of one real equation and one complex equation
that involve only the real unknown $\cg$ and the complex unknown $G$. In accordance with our discussion in Section 4, the real equation contains $\cg$ as an overall factor. Following the steps indicated in Section 4 we divide this factor out and set $\cg=0$, thereby obtainig one equation for $G$ and a constraint. Eliminating $G$, we find the constraint 
\beq\label{diracconstraint}
\mu^2 (4\mu^2+x^2) [(\mu^2-y^2)^2 x^2 + 4\mu^2 y^4 + (1+4\mu^2-8\mu^4) y^2 + 4\mu^4 (\mu^2-1)]=0\,.
\eeq  
Solving (\ref{diracconstraint}) for $x^2$ we obtain 
\beq\label{diracboundary}
x^2 = {1\over (\mu^2-y^2)^2} \left[-4\mu^2y^4 + (8\mu^4-4\mu^2-1) y^2 -4\mu^4 (\mu^2-1)\right]\,,
\eeq
in accordance with \cite{stephanov}.

\pagebreak

\section{Energy Level Dynamics of Non-Hermitean Hamiltonians}

\setcounter{equation}{0}

In much of this paper, we have discussed non-hermitean random
matrices without any spatial structure. Physical problems typically involve spatial structures and thus random
matrices with many zero entries, with non-zero entries representing nearest
neighbor hopping for instance. For such problems, even in the hermitean
case, there is typically no simple analytic approach available and these
problems has to be studied on a case by case basis. Here we will use the
problem of driven vortex lines as an illustrative context within which to
make some general qualitative comments. In the simplest model\cite{nelson}
for this problem, the Hamiltonian $H$ is the sum of a deterministic but
non-hermitean hopping matrix $H_0$ and a random but hermitean matrix
representing pinning by impurities $V$. We have in the simplest
one-dimensional case with $N$ sites
\beq\label{h0}
H_{0 ij} = e^h~\delta_{i+1, j} + e^{-h}~\delta_{i, j+1}\quad , \quad i,j = 1, \cdots N
\eeq
describing a particle which tends to hop in one direction more than in
the opposite direction. The
eigenvalues of $H_0$ (assuming periodic boundary conditions)
can be easily worked out by standard methods to be
\beq\label{spectrum}
E_n = 2~{\rm cos}~( {2\pi n\over N} - ih)\, \quad (n = 0, 1, \cdots, N-1)\,, 
\eeq
thus tracing out an ellipse. (The limits of $h \rightarrow 0$ and
$h \rightarrow \infty$ in which the ellipse collapses to a line segment
on the real axis and expands to a circle are both easily understood.) The
precise form of $V$ reflects what one believes to be an appropriate
representation of the microscopic physics, of course. Some choices that
come to mind are a random nearest hopping Hamiltonian
$V_{ij} = v_i~\delta_{i+1, j} + v_j^*~\delta_{i, j+1}$, or a diagonal
Hamiltonian describing site impurities $V_{ij} = v_i~\delta_{ij}$, or a purely random hermitean matrix $V$. This last case may be most amenable to analytic study, 
but represents the somewhat unphysical situation of being able to hop from 
any site to any other site (the ``infinite dimension" limit). 

Let us begin our discussion of random non-hermitean hopping by making
some simple observation based on symmetry. For $H_0$ real, the non-real
eigenvalues come in complex conjugate pairs. This is true in particular for
$H_0$ in (\ref{h0}), namely $E_n$ and $E_{N-n}$ in 
(\ref{spectrum}) form such a pair. Furthermore, if $\psi$ is an eigenvector (say, on the right) of $H_0$, corresponding to an eigenvalue $E$, $\psi^*$ is an eigenvector (on the right) corresponding to an eigenvalue $E^*$.  For the specific $H_0$ in (\ref{h0}), if $N$ is even, then the eigenvalues also come 
in opposite pairs: $E_n = - E_{n+N/2}$. This follows for a large class of hopping Hamiltonians with ``bipartite symmetry" under an operation in 
which we multiply the basis vectors associated with the odd sites by $(-1)$. 

As we crank up the hermitean $V$ we expect the eigenvalues
tracing out the ellipse to slowly migrate to the real axis, and becoming
real in the infinite $V$ limit. In Figure (1) we present numerical results
for the spectrum of $H_0 + V$ (with periodic boundary conditions), where
$V$ describes random site energies. These site energies are taken from a flat
distribution, symmetric about the origin, and their range is comparable with
the size of the ellipse. We do not bother to display here how the distortion 
of the ellipse varies with the range of the random site energies. 
\begin{figure}
\epsfysize=4 truein
\epsfxsize=4 truein
\centerline{\epsffile{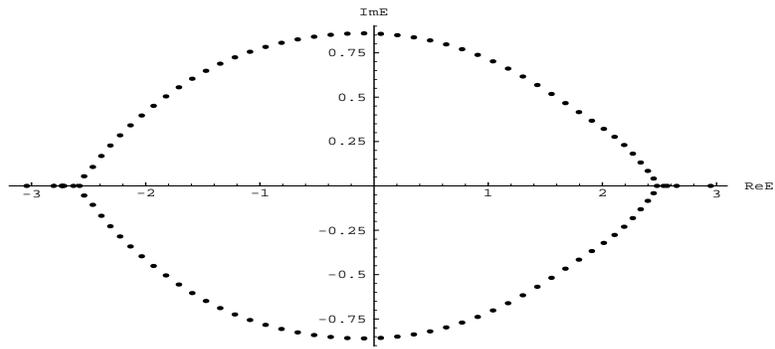}}
\vspace{-0.1 truein}
\caption[]{ The spectrum of (\ref{h0}) with random site energies (with 100
sites). Here $h=0.5$ and the random site energies $|v_i|\leq 1.5$ are drawn from a flat distribution, symmetric about the origin. There are 12 real eigenvalues in all.}
\end{figure}

To the initial surprise of the authors, already for a finite
amount of randomness, there are pairs of eigenvalues which have ``snapped"
onto the real axis. In fact, this ``snapping phenomenon" is easy to
understand qualitatively.

Repulsion of nearby energy eigenvalues is of course one of the fundamental
features of quantum mechanics, and is explained in any elementary texts as
a property of $2\times 2$ hermitean matrices. If we have a pair of eigenvalues
$\pm \epsilon$, and turn on an off-diagonal perturbation $v$ between them,
the eigenvalues become $\pm \sqrt{\epsilon^2+v^2}$, thus moving apart. The
behavior of random hermitean matrices can be understood qualitatively in
this way, and is formalized in the concept of the Dyson gas with its
molecular repulsion.

Now if we have a pair of complex eigenvalues separated in the imaginary
direction, that is, if we have a pair of eigenvalues $\pm i\epsilon$, we
see that under the (hermitean) perturbation $v$ they become $\pm i
\sqrt{\epsilon^2-v^2}$. Our standard intuition fails. Two nearby imaginary eigenvalues attract! Furthermore, they do not migrate gradually to the real axis. Rather, as soon as $v$ becomes larger than $\epsilon$ in magnitude, the
eigenvalues snap to the real axis.

(We note that in the case of an anti-hermitean perturbation (that is, for
$v$ imaginary) the opposite behavior holds: two nearby eigenvalues
separated along the real direction attract, and two nearby eigenvalues
separated along the imaginary direction repel.)

As in the case of hermitean matrices, we can reaffirm the mutual effects of 
nearby energy levels we observed in the simple $2\times 2$ matrix example
by doing a second order perturbation calculation.
For simplicity, we assume that $H_0$ is non-degenerate. Let us then add a perturbation $V$ to $H_0$. We recall that a non-hermitean Hamiltonian $H_0$ 
with eigenvalues $\{E_n^{(0)}\}$ has a set of eigenvectors on the right $H_0 |\psi_n^{(0)}\rangle = E_n^{(0)} |\psi_n^{(0)}\rangle$, as well as a set of eigenvectors 
on the left $\langle \phi_n^{(0)} | H_0  = \langle \phi_n^{(0)} |  E_n^{(0)}$, that form a complete bi-orthonormal system\cite{morse}. With these facts in mind,  perturbation theory for non-hermitean Hamiltonians appears formally the same 
as in the purely hermitean case (as long as one remembers to make the
distinction between eigenvectors on the right and eigenvectors on the left.)
In particular, the correction of $E_n^{(0)}$ to first order in $V$
is simply $\delta^{(1)}~E_n = \langle \phi_n^{(0)} | V | \psi_n^{(0)}\rangle $, and the correction to second order in $V$ is 
\beq\label{perturbation}
\delta^{(2)}~E_n = \sum_{k\neq n} {\langle \phi_n^{(0)} | V | \psi_k^{(0)}\rangle \langle \phi_k^{(0)} | V | \psi_n^{(0)} \rangle
\over E_n^{(0)} - E_k^{(0)}}\,.
\eeq
For a diagonal $H_0$, the bi-orthogonal sets coincide $\psi_n^{(0)}=\phi_n^{(0)}$, forming a single ordinary orthogonal basis, and (\ref{perturbation}) assumes the same form as in perturbation theory of
hermitean operators. In this case we see immediately that if $V$ is hermitean, the original eigenvalues repel in the real direction and attract in the imaginary direction, whereas if $V$ is anti-hermitean, they do the opposite, as expected. 

To get a qualitative feel, let us analyze the effect of a pair of eigenvalues 
on another nearby pair. We concentrate on the $4\times 4$
piece of the Hamiltonian associated with these eigenvalues, ignoring all the other eigenvalues and take  $H=H_0 + V$ with $H_0 = {\rm diag} (i\epsilon, -i\epsilon, c+ia, c-ia)\,\,;\epsilon, a >0$. For simplicity, we take the
perturbation $V$ hermitean with zero diagonal entries, and all other entries
real and equal to $v$ (this should be enough to capture the qualitative
behavior in a typical case.) Assuming that 
$\epsilon << |c|\sim a$, the pair of eigenvalues of $H$ corresponding to $\pm i\epsilon$ would snap first, and we want to study the effect of the pair at $c\pm i a$ on it. Cranking up (\ref{perturbation}) we thus find that to second order in perturbation theory, the first two eigenvalues are
\beqast
\pm i\epsilon \left( 1 - {v^2\over 2 \epsilon^2}\right)
-v^2\left[{1\over c+i(a-\epsilon)} + {1\over c-i(a+\epsilon)}\right]\,,
\eeqast
which to first order in $\epsilon/c^2$ reads
\beq\label{snap}
\pm i\epsilon \left( 1 - {v^2\over 2 \epsilon^2} -2v^2 {c^2-a^2\over (c^2+a^2)^2}\right) - {2 c v^2\over c^2 + a^2} + {\cal O} \left(v^2\epsilon^2/a^3\right)\,.
\eeq
We readily indentify the piece $\pm i\epsilon \left( 1-{v^2\over 2 \epsilon^2}\right)$ in (\ref{snap}) as the first two terms in the expansion of $\pm i\sqrt{\epsilon^2-v^2}$ we mentioned earlier, namely, the exact
value of these eigenvalues when $c\pm ia\rightarrow\infty$.
Thus, under the influence of $V$, $\pm i\epsilon$ indeed attract each other.
Note that to the order to which (\ref{snap}) holds, the pair $c\pm ia$ slows
or accelerates the snapping of the first pair onto the real axis
depending on whether $c^2<a^2$ or $c^2>a^2$. The first pair is also 
repelled by the pair $c\pm ia$ along the real axis. If $c>0$ it moves
to the left, and if $c<0$ it moves to the right.

The elementary discussion just concluded explains qualitatively
the distortion of the ellipse corresponding to $H_0$ that we see in Figure (1).  The eigenvalues near the tips of the ellipse are the first to snap to the real axis, for the obvious reason that they are closer to the real axis. They are then repelled away from the ellipse by the eigenvalues that remain off the real axis. Once on the real axis, they repel the eigenvalues that snapped earlier
further along the real axis.

For random hermitean hopping Hamiltonians, the lore is that the states in
the middle of the band (or possibly only one state) are extended, while the
states on the two ``wings" of the band are localized. For the
one-dimensional case, this can be shown quite explictly by analytical
methods\cite{alsatians}, and for the two-dimensional case,  this can be
studied by renormalization group arguments and verified numerically. In the
present one dimensional context, we might thus expect (or conjecture) that 
the eigenvalues that have snapped onto the real axis correspond to localized states, while the eigenvalues that remain complex correspond to extended states \cite{nelson}. In Figures (2) and (3) we present numerical studies of the participation ratio that support this conjecture or expectation\footnote{Out of the $12$ real eigenvalues we found numerically in this case, we chose to depict the most localized state. Some of the other states corresponding to real eigenvalues may break into 2 or 3 separated lumps, but their participation ratios remain of the same order of magnitude as of the state presented on Fig. (2).}. 
\begin{figure}
\epsfysize=4 truein
\epsfxsize=4 truein
\centerline{\epsffile{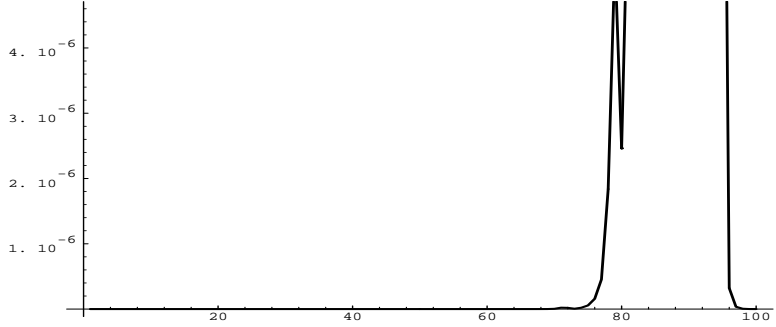}}
\vspace{-0.1 truein}
\caption[]{ The eigenvector $\psi_i$ associated with one of the eigenvalues
(2.9) on Fig. (1) that snapped onto the real axis. The corresponding normalized participation ratio is 0.04\,.}
\end{figure}
The normalized participation ratio for a state $\psi_i$ is defined as 
\beq\label{participation}
P={1\over N} {\sum_{i=1}^N |\psi_i|^2 \over \sum_{i=1}^N |\psi_i|^4}
\eeq
and thus should tend to $1$ for extended states and to $0$ for localized 
states. 
\begin{figure}
\epsfysize=4 truein
\epsfxsize=4 truein
\centerline{\epsffile{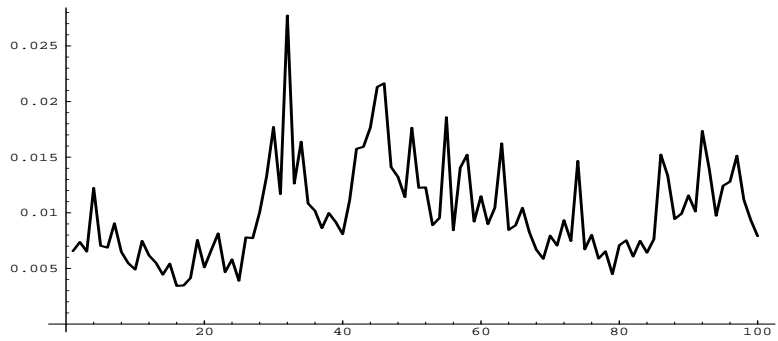}}
\vspace{-0.1 truein}
\caption[]{ The eigenvector $\psi_i$ associated with one of the complex eigenvalues (0.2 + 0.8 i) on Fig. (1) . The corresponding 
normalised participation ratio is 0.8\,. }
\end{figure}

It is worth emphasizing that the method of hermitization given in Section 3 
is independent of the spatial character of the random non-hermitean Hamiltonian. Thus, in principle, it may be used to reduce the problem
of random non-hermitean hopping to the auxilary hermitean problem. 
The deterministic part of the hermitean problem can be interpreted as a 
nearest neighbor hopping of a particle with a binary internal state 
(call it an ``up quark" or a ``down quark"), such that whenever the particle hops it flips its internal state. In any case, it still bears the same spatial 
structure as in the original non-hermitean problem, thus retaining all the difficulties associated with random hermitean matrices with spatial structure.

\pagebreak


\vskip 10mm
\begin{center}
{\bf ACKNOWLEDGEMENTS}
\end{center}
A.Z. would like to thank David Nelson for stimulating his interest in non-hermitean random matrices. He also thanks Edouard Br\'ezin for extensive discussions and the \'Ecole Normale Sup\'erieure for its hospitality. Some of 
the calculations reported here were first done with the help of Edouard Br\'ezin. A.Z. also thanks Freeman Dyson for helpful conversations and the Institute for Advanced Study for a Dyson Distinguished Visiting Professorship.
J.F. would like to thank Gerald Dunne for some references concerning quaternionic analysis. Both authors are obliged to Mark Srednicki for help 
with the numerical calculations. This work was supported in part by the 
National Science Foundation under Grant No. PHY89-04035, and by the Dyson Visiting Professor Funds.


\newpage
\setcounter{equation}{0}
\renewcommand{\theequation}{A.\arabic{equation}}
{\bf Appendix : {A $2\times2$ example}}
\vskip 5mm

In this Appendix we will perform an exceedingly elementary but instructive
exercise. Consider the $2\times 2$ \nh
\beqast
\phi=\left(\begin{array}{cc} 0~~~ & C+i\mu\\{} & {}\\ C^*+i\mu & 0\end{array}\right)
\eeqast
where $C$ is a Gaussian random complex variable and $\mu$ is a real parameter. This is thus the
$N=1$ limit of the problem considered by Stephanov\cite{stephanov}, which we
discussed in sub-section 5.2. The
eigenvalues of $\phi$ are $\lambda=\pm [(C+i\mu)(C^*+i\mu)]^{1/2}$.
Writing $C=r e^{i\theta}$ we have by definition the density of
eigenvalues
\beqra
\rho(x,y)=\int\int dr^2 ~{d\theta\over 2\pi} e^{-r^2} \delta(x-\Re \lambda)~
\delta(y-\Im \lambda)
\eeqra
The evaluation of this integral is elementary and gives
\beq\label{rhoh}\rho(x,y)={4\over \pi}{(x^2+y^2)\over \sqrt{(\mu^2+x^2)(\mu^2-y^2)}}e^{-(x^2-y^2+\mu^2)}~
\theta(\mu^2-y^2)\,.
\eeq
As is indicated by the $\theta$ function, the density of eigenvalues is
non-zero over an infinitely long strip with width $2\mu$. It decreases
exponentially as $x\rightarrow \pm \infty$ and blows up as an inverse
square root as $y \rightarrow \pm \mu$.
Notice that if we impose the condition that $H$ be normal (but non-hermitean), that is $[\phi,\phi^{\dagger}]=0$, $C$ is required to be real, and the density of
eigenvalues collapses to
\beq\label{normal}
\rho(x,y)={1\over 2}P(x)~[\delta(y-\mu) +\delta(y+\mu)]
\eeq
where $P(C)$ denotes the (normalized) probability of distribution of $C$, whatever one might choose. The infinitely long strip collapses into two lines.

\end{document}